\documentclass[aps,twocolumn,pra,showpacs,floatfix]{revtex4}
\usepackage{epsfig}
\usepackage{graphicx}
\usepackage{dcolumn}
\usepackage{amssymb,amsmath}
\usepackage{mathrsfs}
\begin{document}

\title{Fast configuration-interaction calculations for nobelium and ytterbium}

\author{V. A. Dzuba and V. V. Flambaum}

\affiliation{School of Physics, University of New South Wales, Sydney 2052, Australia}

\author{M. G. Kozlov}

\affiliation{Petersburg Nuclear Physics Institute of NRC ``Kurchatov
Institute'', Gatchina 188300, Russia}

\affiliation{St.~Petersburg Electrotechnical University
``LETI'', Prof. Popov Str. 5, 197376 St.~Petersburg}

\begin{abstract}
We calculate excitation energies  for low states of nobelium, including states with open $5f$ subshell. An efficient version of 
the many-electron configuration-interaction method for treating the atom as a sixteen external electrons system has been 
developed and used. The method is tested on  calculations for ytterbium which has external electron structure similar to nobelium.
The results for nobelium are important for prediction of its spectrum and for interpretation of recent measurements. 
Ytterbium is mostly used to study the features of the method.
\end{abstract}

\pacs{31.15.A-,11.30.Er}

\maketitle


\section{Introduction}

Configuration interaction (CI) method \cite{Grant70,KT87} is one of few tools used to calculate electron structure of open-shell many-electron atoms. However, due to huge increase of the computational cost with the number of external electrons, practical application is usually limited to systems with only few (no more than four) external electrons above closed shells. There are no other {\em ab initio} methods to deal with more complicated polyvalent systems. On the other hand, the use of different semi-empirical approaches is questionable when experimental data is poor or absent. Superheavy elements ($Z>100$) \cite{Scha06,EFK15} and highly charged ions \cite{KSCS18} are just two good examples of such systems.
Lack of good theoretical approaches is a big obstacle in the use of complicated atomic systems in fundamental research. An important step to address the problem was done in recent work~\cite{DBHF17}. It was demonstrated that neglecting off-diagonal matrix elements in the CI matrix between highly excited states can be used to reduce the CI problem to a matrix eigenvalue problem with relatively small matrix with modified (compared to standard CI approach) matrix elements. Since the corrections to matrix elements were similar to the second-order perturbation theory corrections to the energy, the method  was called CI with perturbation theory (CIPT). Similar approaches were later used in a number of works~\cite{JGRR17,GCKB18,KaBe18}, while a somewhat different variant of CI+PT method was developed in \cite{RKP01e,ImaKoz18}. This made it possible to perform calculations for such complicated atomic systems as Yb (including states with excitations from $4f$ subshell)~\cite{DBHF17,DFS18}, W~\cite{DBHF17}, Ta, Db~\cite{LDF18a}, Og~\cite{LDF18}, etc.

In this work we further develop the method to make it substantially more efficient. We demonstrate that neglecting the difference between energies of the states of the same excited configuration allows one to separate summation over projections of the total angular momentum of single-electron states from summation over other quantum numbers. Since summation over projections is the same for all similar configurations it can be performed only once and then reused for other similar configurations. This reduces computational time for Yb more than twenty times while the effect on the accuracy of the calculations is negligible. We use Yb atom as an example and then apply the method to nobelium. This allows us to predict No spectrum including states with excitations from the $5f$ subshell. It is also important, that we provide the proof of validity of previous calculations used for interpretation of the experimental measurements. The energy, hyperfine structure and isotope shift for the $^1$P$_1^{\rm o}$ state of several No isotopes were measured \cite{RABBB18} and used together with atomic calculations to extract nuclear parameters of these isotopes \cite{PSSDF18}. Nobelium atom was treated in the calculations as a system with two valence electrons above closed shells. It is known that similar treatment of the $^1$P$_1^{\rm o}$ state of Yb gives very poor results due to the mixing with a close state containing excitation from the $4f$ subshell. This mixing cannot be properly accounted for in the two-valence-electrons-above-closed-shells calculations. We demonstrate that the situation in No is different and corresponding mixing is small. Therefore, interpretation of the measurements based on the two-valence-electron calculations is correct. New energy levels for low states of No including those with open $5f$ subshell have been calculated.

\section{Fast configuration interaction method}

Fast configuration interaction method (FCI) is a modification of the CIPT (configuration interaction with perturbation theory) method introduced in Ref.~\cite{DBHF17}. We start from its brief description using ytterbium atom as an example. We consider Yb as a system with sixteen electrons above closed shells. The CI Hamiltonian has the form
\begin{equation}
H^{\rm CI} = \sum_i^{16} h_i + \sum_{i<j}^{16}\frac{e^2}{r_{ij}}, \label{eq:HCI}
\end{equation}
where $h_i$ is the single-electron part of the Hamiltonian,
\begin{equation}
h_i = c{\mathbf \alpha}\cdot{\mathbf p}_i + (\beta - 1)mc^2 + V^{\rm HF}(r_i).
\label{eq:h1}
\end{equation}
$V^{\rm HF}(r)$ is the self-consistent Hartree-Fock (HF) potential (including nuclear part) obtained in the $V^{N-1}$
approximation, with one $6s$ electron removed from the HF calculations.
The many-electron wave function for sixteen external electrons has the form of expansion over single-determinant
basis states:
\begin{equation}
\Psi(r_1,\dots,r_{16}) = \sum_i c_{i} \Phi_i(r_1,\dots,r_{16}).
\label{e:Psi}
\end{equation}
The basis
states $\Phi_i(r_1,\dots,r_{N_e})$ are obtained by distributing sixteen electrons over
a fixed set of single-electron orbitals.
The coefficients of expansion $c_{i}$ and corresponding energies $E$ are found by solving
the CI matrix eigenvalue problem
\begin{equation}
(H^{\rm CI} - EI)X=0,
\label{e:M}
\end{equation}
where $I$ is the unit matrix, the vector $X = \{c_1, \dots, c_{N_s}\}$, and $N_s$ is the number of many-electron basis states. It is assumed that a few first terms in the expansion (\ref{e:Psi}) represent a good approximation for the state of interest and the rest of the sum is just a small correction. Then one can neglect the off-diagonal matrix elements between the states from the second part of the expansion and reduce the CI problem to one with the small matrix and the modified matrix elements (see Ref.~\cite{DBHF17} for details)
\begin{equation}
\langle i|H^{\rm CI}|j\rangle \rightarrow \langle i|H^{\rm CI}|j\rangle +
\sum_k \frac{\langle i|H^{\rm CI}|k\rangle\langle k|H^{\rm
    CI}|j\rangle}{E - E_k}.
\label{e:PT}
\end{equation}
Here $|i\rangle \equiv \Phi_i(r_1,\dots,r_{16})$, $E_k = \langle k|H^{\rm CI}|k\rangle$, and $E$ is the energy of the state of interest (the same as $E$ in (\ref{e:M})). Since this energy is not known in advance one needs to perform iterations over it.

\begin{figure}[tb]
\epsfig{figure=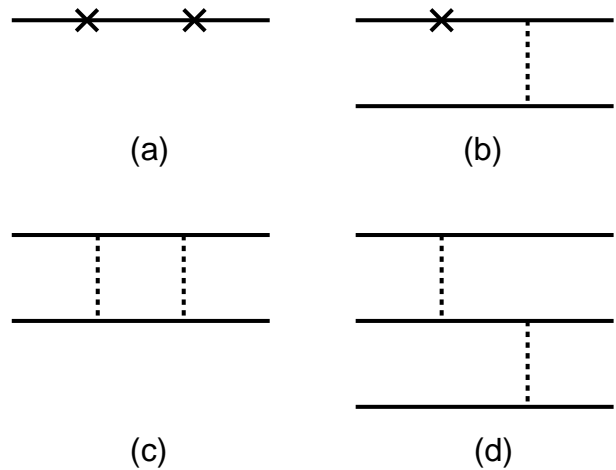,scale=0.9}
\caption{Four diagrams corresponding to three terms in (\ref{e:coef}). Last term in (\ref{e:coef}) corresponds to
diagrams (c) and (d).}
\label{f:diag}
\end{figure}

Starting from this point we consider further modifications to Eq.~(\ref{e:PT}) which lead to the FCI method. Summation in (\ref{e:PT}) goes over all states of excited configurations. If we neglect energy difference between states of the same configuration the summation in (\ref{e:PT}) can be divided in two parts
\begin{eqnarray}
&&\sum_k \frac{\langle i|H^{\rm CI}|k\rangle\langle k|H^{\rm
    CI}|j\rangle}{E - E_k} \approx \nonumber \\
&&    \sum_c \frac{1}{E - E_c} \sum_{k_c} \langle i|H^{\rm CI}|k_c\rangle\langle k_c|H^{\rm CI}|j\rangle .
\label{e:PTm}
\end{eqnarray}
First summation is over excited configurations and $E_c$ is average energy of each configuration. These energies can be expressed analytically in terms of the radial integrals of the Hamiltonian \eqref{eq:HCI} \cite{Grant70,LinRos75,TZSP18}, or calculated numerically. Second summation is over many-electron configuration state functions of a given configuration. These functions differ by the values of projections of the total angular momenta of individual electronic states, but have all other quantum numbers fixed.
Therefore, second summation can be rewritten as:
\begin{multline} \label{e:coef}
\sum_{k_c} \langle i|H^{\rm CI}|k_c\rangle\langle k_c|H^{\rm CI}|j\rangle =
\\=
\sum_{\alpha\alpha'} r_{ij}^{\alpha\alpha'} h_\alpha h_{\alpha'}
+ \sum_{\alpha\beta} s_{ij}^{\alpha\beta}h_{\alpha}q_{\beta}
+\sum_{\beta\beta'}t_{ij}^{\beta\beta'}
q_{\beta}q_{\beta'}\,.
\end{multline}
Here $h_\alpha$ and $q_{\beta}$ are one- and two-electron radial integrals of the CI Hamiltonian (\ref{eq:HCI}), $\alpha$ and $\beta$ 
are the short notations for corresponding sets of electronic quantum numbers. For example, $\alpha = n_1l_1j_1;n_2l_2j_2$. 
Three terms in (\ref{e:coef}) correspond to four diagrams on Fig.~\ref{f:diag} (there are two diagrams for the last term).
Coefficients $r_{ij}^{\alpha\alpha'}$, $s_{ij}^{\alpha\beta}$, and $t_{ij}^{\beta\beta'}$ do not depend on principal quantum numbers of 
one-electron states. They can be calculated only once for a whole set of configurations, which differ by the principle quantum numbers. 
For example, these coefficients are the same for all configurations of the type $4f^{14}nsn's$ ($6<n<n' \leq n_{max}$). 
Since the number of similar configurations can get over a hundred, the effect of reuse of the coefficients is substantial.

\begin{table*}
\caption{\label{t:yb} Comparison of energies (in cm$^{-1}$) and computational times (in minutes)
for low states of ytterbium calculated with the use of the CIPT and FCI methods. Note that all states of the same parity
and total angular momentum are calculated in one run of the program. Therefore, computational time is shown
only for the lowest state. $\Delta$ is the difference
between experimental and theoretical energies; $\Delta_{th}$ is the difference between CIPT and FCI energies.
All theoretical energies are presented with respect to the FCI ground state.
Gain is the ratio of the CIPT computational time to the FCI computational time.}
\begin{ruledtabular}
\begin{tabular}{ll rrr c rrr cc}
\multicolumn{2}{c}{State}&
\multicolumn{1}{c}{Expt.}&
\multicolumn{3}{c}{CIPT} &
\multicolumn{4}{c}{FCI} &
\multicolumn{1}{c}{Gain} \\

&&\multicolumn{1}{c}{\cite{NIST}}&
\multicolumn{1}{c}{Energy} &
\multicolumn{1}{c}{$\Delta$} &
\multicolumn{1}{c}{time} &
\multicolumn{1}{c}{Energy} &
\multicolumn{1}{c}{$\Delta$} &
\multicolumn{1}{c}{$\Delta_{th}$} &
\multicolumn{1}{c}{time} & \\

&&\multicolumn{3}{c}{(cm$^{-1}$)} &
\multicolumn{1}{c}{(m)} &
\multicolumn{3}{c}{(cm$^{-1}$)} &
\multicolumn{1}{c}{(m)} & \\
\hline
\multicolumn{11}{c}{Even} \\
$4f^{14}6s^2$  &  $^1$S$_0$ &    0        &      73    &  73 & 2 & 0 & 0 & 73 &$<1$ &$\sim$6\\
$4f^{14}5d6s$  &  $^3$D$_1$ & 24489  &  27692 & -3203 &68 &27622&-3133&70&3&23 \\
$4f^{14}5d6s$  &  $^3$D$_2$ & 24751  & 27753  &  -3002 &78&27632&-2881&121&3.3&23\\
$4f^{14}5d6s$  &  $^3$D$_3$ & 25271  &  27873  & -2602  & 55  &27812&-2541& 61&2.5& 22\\
$4f^{14}5d6s$  &  $^1$D$_2$ & 27678  & 28244  &  -566  &      &28125&-447&119&&\\

\multicolumn{11}{c}{Odd} \\

$4f^{14}6s6p$  & $^3$P$^{\rm o}_0$ & 17288 & 17870 &-582  &215&17820&-532&50& 9 &24 \\
$4f^{14}6s6p$  & $^3$P$^{\rm o}_1$ & 17992 & 18374 &-382  &267&18264&-272&110&10&27\\
$4f^{14}6s6p$  &  $^3$P$^{\rm o}_2$& 19710 &  20076 & -366 &299&20049&-339&27&13& 23\\

$ 4f^{13}5d6s^2$ &  (7/2,3/2)$^{\rm o}_2$ &  23188  & 24904 &-1716  && 24806&-1618&98&&\\
$ 4f^{13}5d6s^2$ &  (7/2,3/2)$^{\rm o}_3$ &  27445  &  27261 &184  &221&27064&381&197&11& 20\\
$4f^{14}6s6p$ &  $^1$P$^{\rm o}_1$ &  25068 & 24433 & 635 &&24316&752&117& &\\
$4f^{13}5d6s^2$& (7/2,5/2)$^{\rm o}_1$ &  28857 & 29512 & -655 &&29380&-523&132&&\\

\end{tabular}
\end{ruledtabular}
\end{table*}

The results of calculation for energy levels of Yb are shown in Table~\ref{t:yb}. All results are obtained by the same computer code which has options to run in either CIPT or FCI mode. Therefore, the difference between CIPT and FCI results ($\Delta_{th}$) is only due to the neglecting the energy difference between states within the same excited configuration. Some difference between the present and previously published CIPT results~\cite{DBHF17,DFS18} is due to the differences in the size of the set of configurations. Present code uses different algorithm to generate excited configurations from the reference configurations. The data in Table~\ref{t:yb} shows that switching from the CIPT to FCI approaches brings substantial gain in efficiency (more than twenty times for Yb) while having only negligible effect on the energies. Note also that the use of the FCI instead of CIPT method does not affect the calculation of the matrix elements. The form of the calculated wave function is the same in both methods. Calculation of the matrix elements was considered in \cite{DFS18,LDF18a}.

\section{Nobelium}

\begin{table}
\caption{\label{t:no}
Calculated excitation energies (in cm$^{-1}$) and $g$-factors for the lowest states of nobelium.
Comparison with earlier calculations and experiment.}
\begin{ruledtabular}
\begin{tabular}{ll rrrr}
\multicolumn{2}{c}{State} &
\multicolumn{2}{c}{FCI (this work)} &
\multicolumn{2}{c}{Other~\cite{PSSDF18}} \\
&&    \multicolumn{1}{c}{Energy} &
\multicolumn{1}{c}{$g$-factor} &
    \multicolumn{1}{c}{Cut CI} &
      \multicolumn{1}{c}{CI+all} \\
        \hline
$5f^{14}7s^{2}$ & $^1$S$_0$ &        0 &  0      &    0 &\\
$5f^{14}7s7p$  & $^3$P$_0^{\rm o}$ &    20091 &  0.0000 & 16360 & 19567 \\
              & $^3$P$_1^{\rm o}$ &    21201 &  1.4581 & 18138 & 21042 \\
              & $^3$P$_2^{\rm o}$ &    26177 &  1.5000 & 22536 & 26133 \\
              & $^1$P$_1^{\rm o}$ &    29783\footnotemark[1] &  1.0423 & 30237\footnotemark[1] & 30203\footnotemark[1] \\

$5f^{14}7s6d$  & $^3$D$_1$ &    31057 &  0.5000 & 31003 & 28436 \\
              & $^3$D$_2$ &    31132 &  1.1589 & 31223 & 28942 \\
              & $^3$D$_3$ &    31579 &  1.3333 & 31608 & 30183 \\
              & $^1$D$_2$ &    32858 &  1.0078 & 37980 & 33504 \\

$5f^{13}7s^{2}6d$ & $^3$P$^{\rm o}_2$ &    42756 &  1.4414 &  45720 & \\
                 & $^3$H$_5^{\rm o}$ &    44294 &  1.0235 &  49731 & \\
                 & $^5$F$_3^{\rm o}$ &    45452 &  1.2229 &  52172 & \\
                 & $^3$H$_6^{\rm o}$ &    45742 &  1.1667 &  52415 & \\
                 & $^5$K$_4^{\rm o}$ &    46123 &  1.1192 &  53701 & \\
                 & $^1$D$_2^{\rm o}$ &    46718 &  1.0258 &  54016 & \\
                 & $^5$K$_4^{\rm o}$ &    47713 &  1.1143 &  56597 & \\
                 & $^3$F$_3^{\rm o}$ &    47855 &  1.0807 &  56958 & \\
                 & $^1$P$_1^{\rm o}$ &    47952 &  1.1334 &  55695 & \\
\end{tabular}
\footnotetext[1]{Experimental value is 29962~cm$^{-1}$~\cite{LLBB16}.}
\end{ruledtabular}
\end{table}

Nobelium is the heaviest element ($Z$=102) for which experimental spectroscopic data are available. The frequency of the strong electric dipole transition from the ground state to a state of opposite parity and the first ionization potential have been recently measured~\cite{RABBB18,CABB18}. The measurements \cite{RABBB18} include hyperfine structure and isotope shifts for three nobelium isotopes $^{252}$No, $^{253}$No, and $^{254}$No. The data were used to extract nuclear parameters, such as nuclear radii, magnetic dipole and electric quadrupole moments. This procedure relies on the atomic calculations. In particular, an advanced combination of the CI with coupled-cluster method was used~\cite{RABBB18,PSSDF18}. Nobelium atom has the electron structure similar to that of ytterbium. Its ground-state is [Ra]$5f^{14}7s^2 \ ^1$S$_0$. The state for which the frequencies of the transitions were measured was [Ra]$5f^{14}7s7p \ ^1$P$_1$. The calculations treated nobelium as a system with two valence electrons above closed shells. However, it is not known in advance whether such calculations produce good results for No. Similar calculations for the $4f^{14}6s6p \ ^1$P$_1$ state  of Yb give very poor results for hyperfine structure~\cite{DFS18} and electric dipole transition amplitude from the ground state~\cite{DzuDer10} due to the strong mixing with the close state of the same parity and $J$ but with an excitation from the $4f$ subshell, $4f^{13}5d6s^2$  (7/2,5/2)$^{\rm o}_1$ (last line of Table~\ref{t:yb}). This mixing cannot be included in the two-valence-electron calculations. However, treating Yb as a sixteen electron system with the CIPT method leads to dramatic improvement of the results~\cite{DFS18}. The results of the FCI calculations presented in Table~\ref{t:no} show the potentially trouble-making state of the $5f^{13}7s^{2}6d$ configuration in No (last line of Table~\ref{t:no}) is significantly higher on the energy scale of No than a similar state in Yb. Energy interval in No is five times larger and mixing is small. The mixing in the $^1$P$_1$ state of interest is 98\% to 2\% in No (2\% admixture of the state with excitation from $5f$ subshell) and 75\% to 25\% in Yb. This means that the mixing in No can be neglected and the $7s7p \ ^1$P$_1$ state can be treated as a two-valence-electron state.

Note the good agreement of the FCI energies with the only known experimental value and with sophisticated calculations by the CI+all-order method for the two-valence-electron states above closed-shell core of No. There are two major sources of uncertainty in the FCI calculations. One is neglecting core-valence correlations with core states below $5f$. Another is the perturbative treatment of the excited configurations. Both these effects are treated more accurately in the CI+all-order calculations. Therefore, if the mixing with states containing excitations from the $5f$ subshell can be neglected, the CI+all-order calculations are preferable and probably more accurate. From these calculations we know that when No atom is treated as a two-valence-electron system, about 95\% of the core-valence correlations come from the $5f$ electrons (this is also true for the $4f$ electrons of Yb). These correlations are included in the FCI calculations. This explains good accuracy of the results.

\acknowledgments

This work was funded in part by the Australian Research Council and by Russian Foundation for Basic Research under Grant No.\ 17-02-00216. MGK acknowledges support from the Gordon Godfree Fellowship and thanks the University of New South Wales for hospitality.


\end{document}